# Dynamical Features of Open Star Cluster: DOLIDZE 14

**Gireesh C. Joshi[1] and R.K. Tyagi[2]**

*[1,2]Department of Physics H. N. B. Govt. P. G. College Khatima, U. S. Nagar-262308*
*E-mail: [1]gchandra.2012@rediffmail.com*

**Abstract**—*The radial density profile (RDP) of cluster DOLIDZE 14 provides the radius of 9.8 arcmin, it is based on consideration of first plateau region of RDP. This value is less than radius which comes through consideration of the second plateau region. The stars, which are inside of the cluster radius, are used to estimate the distance-modulus and E(I −R) by fitting theoretical isochrone on the colour-magnitude diagram (CMD). The best fitted stellar isochrone of solar matellacity provides these values as 11.15 mag and 0.25 mag, respectively. These results are used to estimate the mass-function slope of the cluster, which comes to be different that of Salpeter-Value. The enhancement of brighter stars does not found within the cluster region which indicates its old age. The Mass-Segregation phenomena of cluster is found between massive and lighter stars but not found in fainter stars, having magnitude greater than 15 mag in I-band. The relaxation-time is very little in the comparison of its age. The mean proper motion of cluster's radial zones is found to be high in the comparison of field region (12.77-15.18 arcmin). This is indicative of weak-gravitational bond among the stars of the cluster.*

## 1. INTRODUCTION

A group of stars i.e. open star cluster (OSC) is a loosely gravitational bound system[1] and dynamic in nature. These systems are also used for tracing tool of the process of evolution of galactic disc[2]. These systems found within galactic plane, therefore, known as galactic clusters. Their shape and behaviour depend on proper motion and luminosity (brightness) of the individual member. The brightness of members represent by logarithm magnitude scale and further directly related to their masses and also provide the information about the total emitting energy per sec from OSC, which defined by the luminosity function (LF). It varies from the cluster center to circumference. In addition, the LF depends on the stellar density (no. of stars per unit area), which decreases with radial distance. The cluster region divided into core region (nuclei) and coronal region on the basis of stellar density [3]. The region, which starts from the center to a distance where the stellar density, reduce to half of its central value, known as the core region of cluster, whereas, remaining region is the coronal region of the cluster. The initial mass function (IMF) of the cluster is defined by stellar mass per unit volume. The direct measurement of the IMF does not possible due to the dynamical nature of the cluster. The dynamical characteristic of cluster members found in such a way that the massive stars move towards to the core region, whereas, lighter moves towards to the coronal region. This phenomenon defined by the mass-segregation of cluster and in this process, the individual member of OSC exchanges the energy in such a way that their velocity distribution reaches towards the Maxwellian equilibrium. The total time for completing this process is known as the dynamical relaxation time, τ(E), of the cluster. It is firstly introduced as an OSC by Alter (1970)[4]. Joshi et. al (2015)[5] have been derived the basic parameters (distance, age, reddening, mean-proper motion etc.) of DOLIDZE 14. The core-radius and radius of the cluster have been estimated by Joshi et al. (2015) [5] as 2.76±0.81 arcmin and 9.8±0.2 arcmin, respectively. The dynamical study of this cluster is useful to understand the stellar encounters, the behaviour of members' motion and influence of molecular clouds within it. We are used to PPMXL catalogue for studying the dynamical properties of the cluster. The detail procedure of estimation of dynamical properties is described in the present work.

## 2. RADIUS AND CORE-RADIUS

The cluster coordinate is found by Joshi et. al. (2015)[5] as $(4^h : 6^m : 26.6^s, +27^d : 22^m : 26.4^s)$ in RA-DEC plane. We have been estimated the cluster radius in various pass-bands. These results indicate that the cluster radius is varied in various pass-bands. The cluster characteristic is not seen by the data of pass-bands of Infra-red (IR) (results are listed in Table 1). The core-radius of the cluster is estimated through the fitting of King empirical model on the RDP. The stellar density, $(\rho_r)$, at a distance r from the center is given by following expression,

$$\rho_r = \rho_{bg} + \frac{\rho_0}{1 + (\frac{r}{r_c})^2},$$

where $\rho_{bg}$, $\rho_0$ and $r_c$ are the background stellar density, central peak stellar density and core radius of the cluster, respectively. The RDP characteristics of cluster indicate two plateau regions. Joshi et al. (2015)[5] have been estimated the cluster-radius considering first plateau region while same is found through second plateau region at present work. The new core-radius and radius values (through PPMXL catalogue) are found to be 2.99±0.59 arcmin and 11.56±0.24 arcmin, respectively. We have been seen significant change in the cluster radius due to the considering of the second plateau region as a filed region. It is possible that the first plateau region occurred due to the enhancement of those stars, which leave the cluster-gravitational bond and merge with field region. In addition, the second plateau region may be field region where no star comes from the cluster. The cluster stellar density distribution of various pass-bands has been shown in Figure-01. The continuous smooth blue line and dashed red lines in each panel represent the best fit of King Model and the background stellar density, respectively. Once, the stars in first plateau region were members of OSC but now they are not. So, their distribution in CMD may be similar to cluster members, cause no distribution in estimation of age, distance and reddening. On the basis of these facts we concluded that the stars of first plateau region of field region are effectively useful to estimate the exact cluster shape within possible gravitational boundary, however, consideration of second plateau region consists stars of first plateau region with cluster.

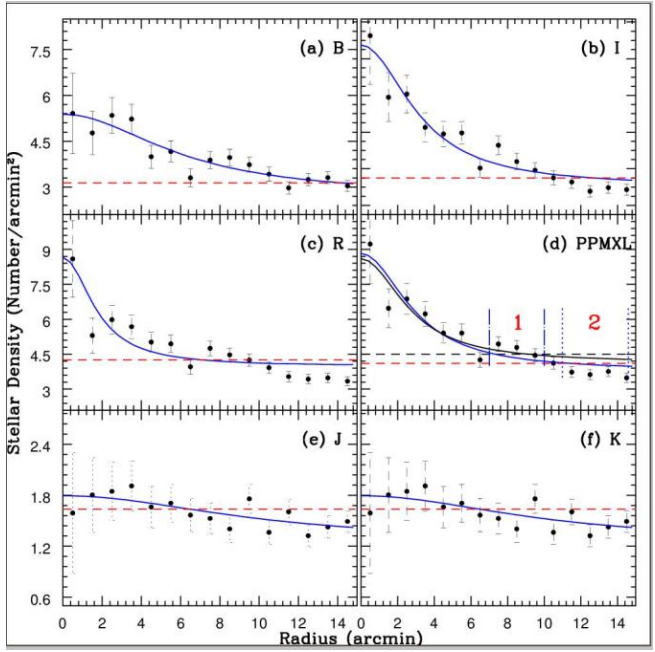

**Fig. 1: The radial density profiles in various passbands.**

Table 1. The mass function values in various magnitude bin.

| Pass band | B | I | R | PPMXL |
|---|---|---|---|---|
| $\rho_0 - \rho_{bg}$ | 2.67±0.39 | 4.62±0.37 | 4.67±0.62 | 3.78±0.33 |
| Core radius (arcmin) | 6.18±1.79 | 3.23±0.56 | 1.77±0.48 | 2.99±0.59 |
| Radius (arcmin) | 14.40±0.20 | 12.89±0.21 | 7.06±0.24 | 11.56±0.24 |

## 3. CMD: DISTANCE AND LOG AGE

The distribution of stars of different radial zones has been shown in Figure-02. In the (d), (e) and (f) panels of this Fig. (stellar distribution of purely field region), we have not seen stellar sequence like as cluster region ((a), (b) and (c) panels) but stellar sequence shows a gap between right and left stellar sequence of field region. This gap arises due to the absence of members of the cluster. The left and right stellar sequence stars may be found for foreground and background field stars with respect to present studied cluster. The distance and age of any cluster are easily estimated by fitting of theoretical isochrone of solar matellacity on colour-magnitude diagram (CMD). The best solution of isochrone gives the apparent distance modulus and log (age) of cluster as 11.15 mag and 9.1, respectively.

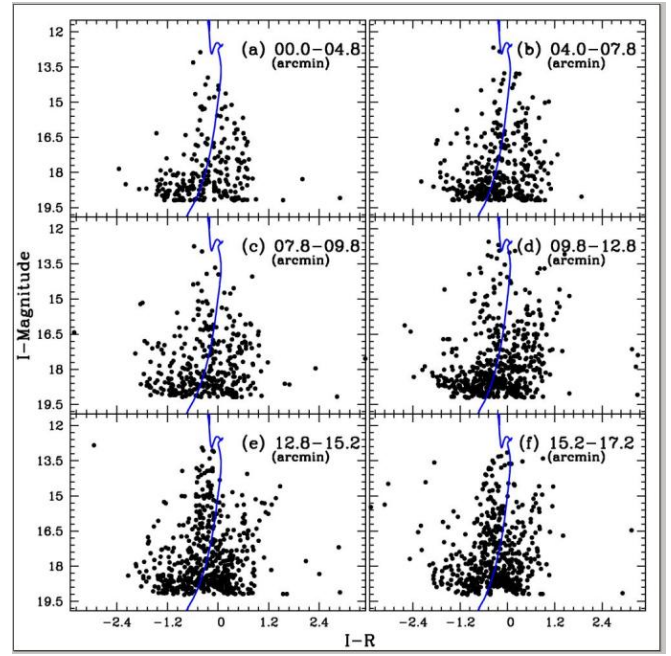

**Fig. 2: This Fig. illustrates the influence of the field stars in various radial zones, the range of each radial zone is written in each panel.**

It is well known fact that the reddening value of any photometric pass-band depends on its effective wavelength and given as $R_i \propto \lambda^{1/2}$. The normal reddening value ($R_V$) is 3.1 for the photometric V-band. We have computed the value of Ri i.e. 2.6 through the following relation,

$$\frac{R_i}{R_V} = \sqrt{\frac{\lambda_i}{\lambda_V}},$$

where $\lambda_i$ and $\lambda_V$ are the effective wavelength in I and V band and their values are 551 nm and 803 nm. The absolute distance modulus value ($M_o$) is found to be 11.15 mag which is very close to 11.12 mag as estimated by Joshi et al. (2015). The distance of a cluster is found to be 1.69 kpc through the formula, $10^{(M_o + 5)/5}$.

## 4. LUMINOSITY FUNCTION (LF)

The stellar distributions in per magnitude bin are defined by the luminosity function (LF). For LF, we have determined the number of stars in a different magnitude bin of unit width in I−band and K-band. We have shown the characteristic LF of cluster region and field region in (a) and (b) panels of Figure-3, in which the blue and red solid curves represent LF characteristic (through the magnitude of stars in I−band and K−band) for both regions respectively. It is mandatory that both regions have equivalent area, therefore, we have chosen a field region equal to the cluster region in the point of view of to see the variation of LF of the cluster with respect to the field region. This Fig. indicates a low stellar enhanced in the lower magnitude bins and high stellar enhanced in the higher magnitude bins of cluster region with a comparison to the field region. The most massive members may not remain so long time due to the cluster old age, therefore, we have not seen any clear stellar enhanced for the massive stars of cluster in comparison of the field region. The enhancement of luminosity function is found only in higher magnitude bins of the cluster. The luminosity function is not itself give any clear picture of dynamical features of members of cluster but it is indicated that it will be well understood by the knowing of features of stellar distribution in different mass range. As a result, the cluster characteristic is found due to the presence of fainter members. The detected stars in nir-IR bands comparatively very low compare them as detected in USNO-B bands.

## 5. MASS FUNCTION

The stellar distribution within mass range is providing the mass function value [6]. The IMF determines the subsequent evolution of the cluster with the star formation rate [7]. The dynamical feature of the cluster does not provide opportunities to direct estimation of the IMF, therefore, mass function (relative number of stars per unit mass) has determined. If, there are N stars are present in the mass range from $m_1$ to $m_2$ then mass function may be expressed as $\varphi = N/\{\log(m_2)-\log(m_1)\}$[8]. The mass function provides the Salpeter-value (-1.35)[9] by following expression;

$$\Gamma = \frac{log(\Phi)}{log(\bar{m})}.$$

The change in mass function with respect to the average mass range has been shown in (c) and (d) panels of Figure-3 for I−band and K−band, respectively. The mass function values have been listed in Table 2 for different mass range of I−band and K−band. We have found high value of mass function slope values for the cluster in comparison of Salpeter-value. The stellar number is found very low in near−IR band compares to I−band. The mass function values depend on stellar magnitude. It is remarkable results that the mass range in I−band is close to K−band. The mass range of I−band is found to be 0.55 $M_\odot$ to 1.82 $M_\odot$ while this range is found to be 0.60 $M_\odot$ to 2.11 $M_\odot$. The fainter stars of K−band of the cluster shows higher incompleteness. These results indicate the low-mass stars may radiate low energy in near-Infrared band, whereas, massive stars may radiate less energy in I−band (visual wavelength). This result of massive stars may occur due to vanish of hydrogen fuel and may be radiating mostly energy in the IR region. Such type stars should be fainter stars in I−band leads as low-mass stars; however, they have massive stars. These facts are good indications for showing stellar mass ariation with various pass-bands.

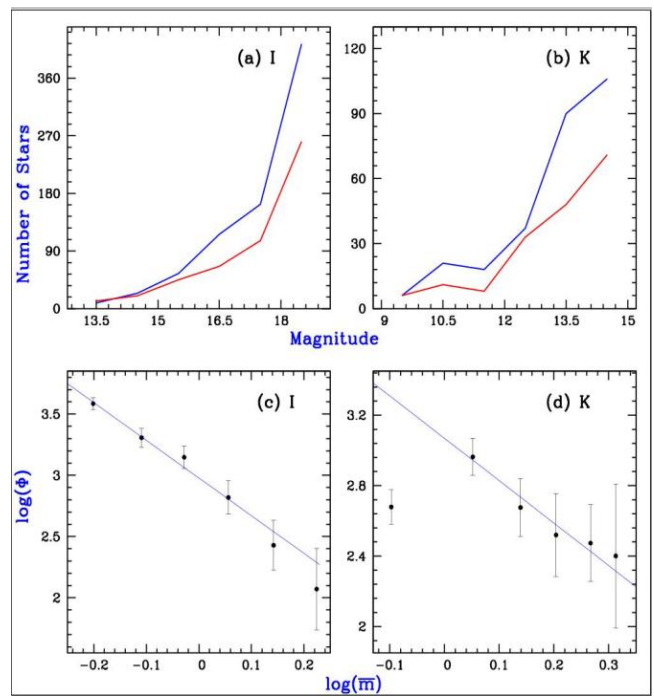

**Fig. 3: (a) and (b) are represented the luminosity function vs magnitude diagram of the cluster in I−band and K−band respectively. (c) and (d) depict the mass function values for the cluster in I−band and K−band, respectively.**

## 6. MASS-SEGREGATION

In the process of mass-segregation, heavy members move towards the core region of the cluster while lighter members move towards the coronal region. It may be a result of equipartition of energy through stellar encounters [10]. To reduce the field star contamination in the cluster, we reject those stars which are far away from the stellar distribution of the main sequence in CMDs plane [11]. We have been divided the cluster members into three groups according to their K−band magnitudes as 8-12 mag, 12-14 mag and 14-16 mag while for I−band magnitude of cluster-members the magnitude range of groups are 13-15 mag, 15-17 mag and 17-19 mag.

The cumulative distribution of stars of these groups is depicted by blue, pink and red solid lines, respectively, as depicted in Figure-4. The cumulative distributions of these groups indicate that the lower mass stars move towards the coronal region while massive stars are moveing towards the coronal region. As a result, the probability of further mass segregation process may be low.

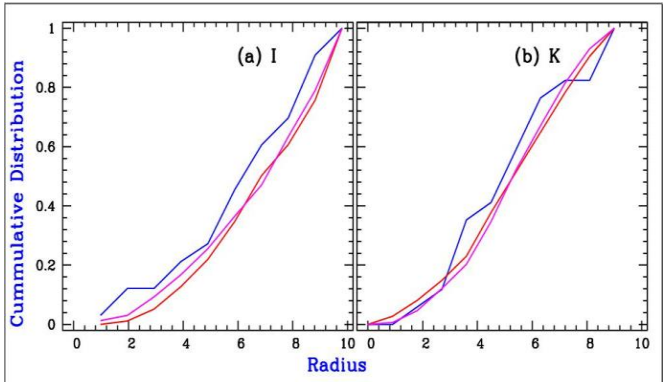

**Fig. 4: The cumulative distribution of stars with cluster radius.**

**Table 2.** The mass function values in various magnitude bins.

| I range (mag) | Mass range $M_\odot$ | $\bar{m}$ $M_\odot$ | $log(\bar{m})$ | N | $log(\Phi)$ | $e_{log(\Phi)}$ |
|---|---|---|---|---|---|---|
| 13-14 | 1.823-1.529 | 1.677 | 0.224 | 9 | 2.071 | 0.333 |
| 14-15 | 1.529-1.245 | 1.387 | 0.142 | 24 | 2.429 | 0.204 |
| 15-16 | 1.245-1.027 | 1.136 | 0.056 | 55 | 2.819 | 0.135 |
| 16-17 | 1.027-0.849 | 0.939 | -0.028 | 116 | 3.147 | 0.093 |
| 17-18 | 0.849-0.706 | 0.778 | -0.109 | 163 | 3.307 | 0.078 |
| 18-19 | 0.706-0.551 | 0.629 | -0.201 | 413 | 3.586 | 0.049 |

| K range (mag) | Mass range $M_\odot$ | $\bar{m}$ $M_\odot$ | $log(\bar{m})$ | N | $log(\Phi)$ | $e_{log(\Phi)}$ |
|---|---|---|---|---|---|---|
| 09-10 | 2.113-2.000 | 2.056 | 0.313 | 6 | 2.400 | 0.408 |
| 10-11 | 2.000-1.700 | 1.850 | 0.267 | 21 | 2.473 | 0.218 |
| 11-12 | 1.700-1.500 | 1.600 | 0.204 | 18 | 2.520 | 0.236 |
| 12-13 | 1.500-1.253 | 1.376 | 0.139 | 37 | 2.675 | 0.164 |
| 13-14 | 1.253-1.000 | 1.126 | 0.052 | 90 | 2.963 | 0.105 |
| 14-15 | 1.000-0.600 | 0.800 | -0.097 | 106 | 2.679 | 0.097 |

## 7. DYNAMICAL RELAXATION-TIME

The time taken for completing the above said phenomena is expressed in the terms of dynamical relaxation-time ($\tau(E)$) which estimated by following expression[12],

$$\tau(E) = \frac{8.9 \times 10^5 \sqrt{NR_h^3}/m}{log(0.4N)},$$

In this expression N stands for no. of MS stars of cluster, $m$ is the average mass of member and $R_h$ stands for the radius of half mass (generally taken as the half of cluster radius). The value of $R_h = \theta \times d$. The value of $R_h$ comes out as 1.6 pc using the relation, taken as in the terms of radian and distance d estimated in pc. In the near-IR's K−band, the total mass of the cluster is found to be 204.93 times of solar mass ($m_\odot$) through 141 stars of the main sequence, which gives average stellar mass (m) about 1.45 $m_\odot$. The value of $\tau(E)$ is computed as 3.58 Myr. Similarly, the cluster mass is found to be 606.37

$m_\odot$ by count masses of 780 stars as detected in I-band of cluster. These results give the values of m and $\tau(E)$ as 0.777 $m_\odot$ and 22.8 Myr. Both estimated values of $\tau(E)$ are less than the cluster age i.e. 1.26±0.04 as estimated by Joshi et. al. (2015)[5]. This fact indicates that the most members are reaching to the state of Maxwellian-velocity, although, the relaxation time in visual band is found to be six times that of finding in the IR band. The variation occurred due to the detection of a larger fraction of fainter stars only in visual band.

## 8. VARIATION OF MEAN PROPER MOTION VALUES

The mean-proper motion of whole cluster has been estimated through the Gaussian analysis. The same number of stars (933 stars) has used to estimate the mean proper motion of the cluster through the Gaussian method as used in iteration method. For this purpose, the number of stars has calculated in each bin of proper motions. The size of each proper-motion bin is 5 mas/yr. The Gaussian fit methods give the mean proper motion of the cluster in RA and DEC directions as −0.74±0.30 mas/yr and −5.28±0.32 mas/yr respectively (as depicted in Figure-5). The proper-motion variation within the cluster radius has used to show the field star fluctuation within the cluster region. For this purpose, the cluster is divided into various rings as mentioned table 1. The mean-proper motion of these rings was calculated by using the following formula,

$$\mu = \sqrt{\mu_x^2 + \mu_{xy}^2},$$

where $\mu_x$ and $\mu_{xy}$ are the mean proper-motion of members of the cluster in the RA and DEC directions respectively. The mean proper motion of the cluster has been found as 5.32±0.43 mas/yr. The mean-proper motion values in different shells have listed in Table 3. The decreasing value of mean proper motion of various radial zone of the cluster (except first one) shows that the field stars have low proper motion values than of cluster members. It is also noticeable that the possible last radial zone (9.8 − 12.8 arcmin) having the mean proper motion similar to that of the field region having range 12.8 − 14.8 arcmin.

**Table 3.** The mean proper motion vlaues of cluster members in different radial zone.

| Radius Range (arcmin) | 00 − 02 | 02 − 04 | 04 − 06 | 06 − 08 | 08 − 9.8 | 9.8 − 12.8 | 12.8 − 14.8 |
|---|---|---|---|---|---|---|---|
| Proper-Motion (RA) ($\mu_x$) | -5.73 | -7.40 | -5.14 | -5.51 | -4.60 | -4.11 | -4.19 |
| Proper-Motion (DEC) ($\mu_{xy}$) | 0.04 | -2.80 | -0.65 | -0.63 | -0.65 | -0.76 | -0.78 |
| Proper Motion ($\mu$) | 5.73 | 7.91 | 5.18 | 5.55 | 4.65 | 4.18 | 4.26 |

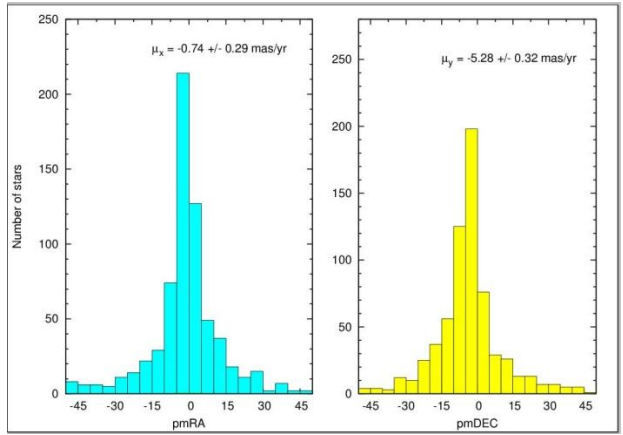

**Fig. 5: The mean proper motion of cluster in RA and DEC directions has estimated through Gaussian-fit method. For this purpose, the histogram of mean proper motion of stars has been represented here.**

## 9. SEARCH OF VARIABLES WITHIN CLUSTER

GCJ has been observed this cluster on the date 13 Oct, 2014 through 1.04m Sampurnanand telescope of ARIES at Manora Peak, Nainital. There are total 52 science frames are taken in I-band duration of 3 hours 15 min. The bias correction and flat fielding have done through bias and flat frames, which were observed on the same date. The exposure time of each frame is 150 sec. We have been using the secondary methodology to reduce atmospheric-effect/ estimation-error. We have detected 200±50 stars in each frame. Since, we have taken first frame as a reference frame, which is not standardized through Landolt field stars. As a result, the secondary standardization methodology provides the apparent magnitude of stars, which are corrected to the reference of the first frame. Due to the lack of the absolute magnitude of stars, we have computed the variation of the magnitude of the stars by transforming their mean apparent magnitude into zero. The variation of magnitude of the stars with respect to time is known as light curve. By deep investigate of light curves of stars, we have found 3 variable stars in observed 15×15 arcmin² field of view of the cluster. This observed region is a small portion of the whole cluster. The coordinates of detecting stars are found in pixel through 2k×2k charge couple device (CCD) camera. These pixel coordinates have been transformed into RA and DEC through astrometry. The 'PerSea.exe' software has been utilized to estimate the period of variable stars. The periodogram of these variables are indicated that there are two regular and one irregular variables. Both regular variables may be δ-Scuti stars on the basis of their period values, whereas, irregular variable may be flare star. The basic information of these variables are listed in Table 4. Their light curves and phase diagrams have been depicted in Fig. 6.

**Table 4.** The detected variable stars of cluster.

| Variable | RA | DEC | Period (days) | Amplitude (mmag) |
|----------|------------|-------------|-----------------|------------------|
| $V_1$ | 4 : 07 : 11.98 | 27 : 15 : 13.3 | 0.0606±0.0068 | 270 |
| $V_2$ | 4 : 06 : 56.83 | 27 : 13 : 09.8 | 0.0714±0.0074 | 474 |
| $V_3$ | 4 : 07 : 02.40 | 27 : 24 : 24.2 | – | 474 |

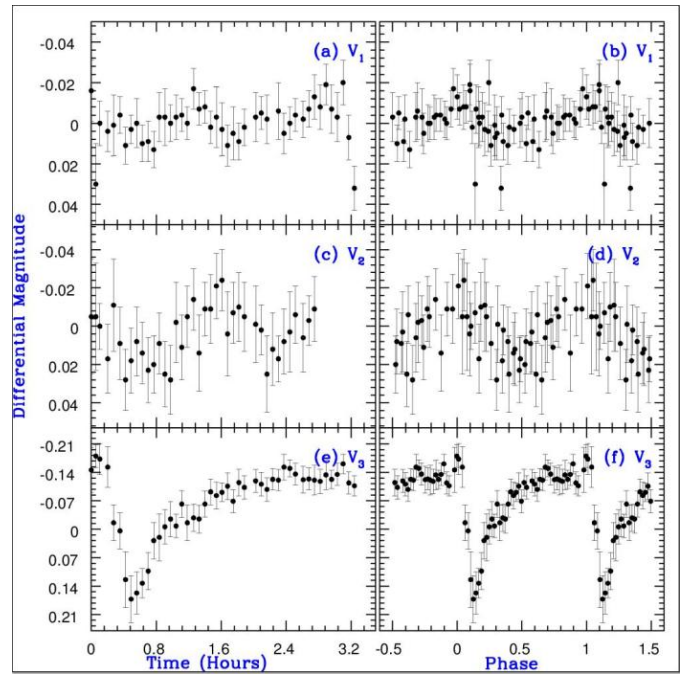

**Fig. 6: The left panels represent the light curves of identified variable stars, whereas, the right panels show their corresponding light-folded-curve or phase diagram.**

## 10. CONCLUSION

The low stellar density is found in B-band though there is no cluster characteristic in infra-red bands. The new core radius is found to be 2.99±0.59 in the case of field region consideration of the second plateau region, which is close to 2.76±0.81 as obtained by Joshi et al.(2015)[5]. The stellar distribution of various radial zones is founded excellent tool to understand the cluster existence. The stellar sequence of cluster is not found after 9.8 arcmin which gives strong supporting fact for cluster radius. The luminosity function is found to be higher with respect to field region in the higher magnitude bin. The slope values of mass function are found to be −3.08±0.22 and −2.41±0.29 for I-band and K-band respectively, which are more numerical value than that of Salpeter-value. The mass-segregation phenomena did not find clearly for cluster in fainter limit of magnitude. Our analysis also indicates that the MF slope value turns towards to a positive value with the increment of incompleteness of data. The values of dynamical relaxation-time and cumulative distribution of stars with radial distance are shown the low probability of mass-segregation phenomena within the cluster. Furthermore, the mean proper motion variation of various radial zones indicates that the mean-proper motion of cluster members is found to be higher than that of field stars. We are identified three variable stars, in which two are δ−Scuti stars and one is flare star.

## 11. ACKNOWLEDGEMENTS

GCJ is thankful to Vivek Joshi (lecturer, P. P. S. V. M. I. College, Nanakmatta) for improving the scientific language of this paper. GCJ is also acknowledgement to ARIES for providing observing facility duration Oct, 2012 to April 2015.

## REFERENCES

[1] Payne-Gaposchin C., "Stars and Clusters", Harvard University Press, Cambridge, 1979.

[2] Frinchaboy M. P. and Majewski R. S., "Open clusters as Galactic Tracers 1. Project Motivation, Cluster Membership and Bulk Three-Dimensional Kinematics", The Astrophysical Journal, 136, 2008, pp. 118-145.

[3] Bukowiecki L., Maciejewski G., Konorski P. and Neidziewa, "Open clusters in 2MASS photometry II. Mass Function and Mass Segregation", Acta Astronomica, 62, 2012, pp. 281-296.

[4] Alter G., Balazs B. and Ruprucht J., "Catalogue of Star Clusters and associations", 1970 and 2nd Ediation, Akademial Kiada Budapest, Printed in Hungary (card form).

[5] Joshi G. C., Joshi Y. C., Joshi S. and Tyagi R. K., "Basic parameters of Open Star ClustersDOLIDZE 14 and NGC 110 in Infrared Range", New Astronomy, 40, 2015, pp. 68-77.

[6] Scalo J., "The Stellar Mass Function", Fund. Cos. Phy., 11, 1986, pp. 1-278.

[7] Kroupa P., "Binary Systems, Star Clusters and Galactic-field Population. Applied Stellar Dynamics", Science, 295, 2002, pp. 82-91.

[8] Kholopov P. N., "The Unity in the Structure of the Star Clusters", Soviat Astronomy, 12, 1969, pp. 625-631.

[9] Salpeter E. E., "The Luminosity Function and Stellar Evolution", The Astrophysical Jounal, 121, 1955, pp. 161-167.

[10] Pandey A. K., Nilakshi, Ogura K., Sager R. and Torasawa K., "NGC 7564: An Interesting Clusters to Study Star Formation History", Astronomy & Astrophysics, 374, 2001, pp. 504-522.

[11] Bonatto Ch., Bica E. and Santos J., "Spatial Dependence of 2MASS Luminosity and Mass Function in the Old Star Cluster NGC 168", Astronomy & Astrophysics, 433, 2005, pp. 917-929.

[12] Jr. L. S. and Hart M. H., "Random Gravitational Encounters Evolution of Spherical Systems I. Method", The Astrophysical Journal, 164, 1971, pp. 399-409.